\documentclass[letterpaper]{article} 
\usepackage{aaai2026}
\usepackage{times}  
\usepackage{helvet}  
\usepackage{courier}  
\usepackage[hyphens]{url}  
\usepackage{graphicx} 
\usepackage{natbib}  
\usepackage{caption} 
\usepackage{algorithm}
\usepackage{algorithmic}
\usepackage{microtype}
\usepackage{multicol}
\usepackage{multirow}
\usepackage{url}
\usepackage{booktabs}
\usepackage{times}
\usepackage{latexsym}
\usepackage{amssymb}
\usepackage{graphicx}
\usepackage{amsmath}
\usepackage{booktabs} 
\usepackage{lineno}
\usepackage{xcolor}
\usepackage{todonotes}
\usepackage{array}
\usepackage{subcaption}
\usepackage[T1]{fontenc}
\usepackage{xspace}
\usepackage{graphicx}
\usepackage{subcaption}
\usepackage[utf8]{inputenc}
\usepackage{tabularx} 
\usepackage{booktabs}
\usepackage{array}
\usepackage{microtype}
\usepackage{amsmath}
\usepackage{amsthm}\usepackage{inconsolata}
\usepackage{enumitem}
\usepackage{xcolor}
\usepackage{ragged2e}
\usepackage{pifont}
\newcommand{\cmark}{\ding{51}}%
\newcommand{\xmark}{\ding{55}}%
\definecolor{mybgcolor}{HTML}{ea998f} 
\definecolor{mybgcolor2}{HTML}{c4d1f7}
\definecolor{mybgcolor3}{HTML}{ffecc1}
\definecolor{mybgcolor4}{HTML}{aec5a8}

\definecolor{darkblue}{rgb}{0, 0, 0.5}

\usepackage{newfloat}
\usepackage{listings}
\DeclareCaptionStyle{ruled}{labelfont=normalfont,labelsep=colon,strut=off} 
\floatstyle{ruled}
\newfloat{listing}{tb}{lst}{}
\floatname{listing}{Listing}
\title{ConvMix: A Mixed-Criteria Data Augmentation Framework \\ for Conversational Dense Retrieval}
\author{
    Fengran Mo\textsuperscript{\rm 1}, Jinghan Zhang\textsuperscript{\rm 2}, Yuchen Hui\textsuperscript{\rm 1}, Jia Ao Sun\textsuperscript{\rm 1}, Zhichao Xu\textsuperscript{\rm 3}, Zhan Su\textsuperscript{\rm 1}, Jian-Yun Nie\textsuperscript{\rm 1}\\
}
\affiliations{
    \textsuperscript{\rm 1}University of Montreal, Canada;
    \textsuperscript{\rm 2}Clemson University, USA;
    \textsuperscript{\rm 3}University of Utah, USA \\

    fengran.mo@umontreal.ca
}

\begin{document}

\maketitle

\begin{abstract}
Conversational search aims to satisfy users’ complex information needs via multiple-turn interactions. The key challenge lies in revealing real users’ search intent from the context-dependent queries. Previous studies achieve conversational search by fine-tuning a conversational dense retriever with relevance judgments between pairs of context-dependent queries and documents. However, this training paradigm encounters data scarcity issues.
To this end, we propose \textbf{ConvMix}, a mixed-criteria framework to augment conversational dense retrieval, which covers more aspects than existing data augmentation frameworks.
We design a two-sided relevance judgment augmentation schema in a scalable manner via the aid of large language models.
Besides, we integrate the framework with quality control mechanisms to obtain semantically diverse samples and near-distribution supervisions to combine various annotated data.
Experimental results on five widely used benchmarks show that the conversational dense retriever trained by our ConvMix framework outperforms previous baseline methods, which demonstrates our superior effectiveness.

\end{abstract}

\section{Introduction}
Conversational search aims to satisfy users' complex information needs via multiple-turn interaction in natural languages. The central hurdle lies in accurately understanding users’ genuine search intent in the conversational session, given that their queries are context-dependent and prone to linguistic issues like omission, coreference, and ambiguity~\citep{gao2022neural,mo2024survey}.

When tackling conversational search, an intuitive approach is to use the conversational query rewriting (CQR) technique that breaks down the task by employing a query rewrite model to convert the query of the current turn into a de-contextualized one and then conducting an ad-hoc search retriever with the rewritten query~\cite{elgohary2019can}. 
This approach allows for the use of existing retrievers for the search process. However, the CQR paradigm is challenging to directly optimize the rewriting towards search, due to the training signals do not directly impact the downstream retrieval tasks~\cite{wu2022conqrr,mo2023convgqr,mao2023search,ye2023enhancing,mao2023large,yoon2024ask}. 
Another paradigm is to conduct conversational dense retrieval (CDR) that aims to train a conversational dense retriever to grasp the search intent by implicitly learning the latent representations of encoded queries and documents, which enable the model to naturally learn the relevant patterns between context-dependent queries and documents based on the relevance judgment signals~\cite{yu2021few,kim2022saving,mao2023learning,jin2023instructor,mo2024history,lupart2024disco}. 
Although these approaches achieve recognized improvements in conversational search, they encounter the data scarcity issue~\cite{mao2022convtrans,dai2022painting}. This is because the conversational search engines are not widely deployed, and thus the available user query logs to generate relevance judgments for dense retriever training are not as abundant as the ad-hoc search scenarios~\cite{bajaj2016ms}.

Constructing conversational search datasets manually is a costly and difficult endeavor, so an intuitive alternative to enrich the training data is to automatically augment the existing data.
Some prior studies~\cite{mao2022convtrans,dai2022painting} attempt to generate additional annotated data, but the generation process remains complex and relies on the assumption that a substantial amount of relevant in-domain data is available to build data augmentation frameworks, which is infeasible for scalable data augmentation.
Recent studies~\citep{huang2023converser,mo2024convsdg} aim to augment existing datasets with the aid of large language models (LLMs) for scalable generation, following the principle of reusing existing manually annotated data in conversations.
However, these approaches simply instruct the LLMs for data generation from a single aspect, e.g., providing the few-shot examples, which cannot control and improve the diversification of the generated data. 
\citet{chen2024generalizing} further develops a generalized data augmentation framework for conversational dense retrieval via LLM-cognition with a quality control mechanism to filter out those noisy and intent shift turns.
However, it encounters the data distribution shift issue when combining multiple types of annotated data, i.e., the original human-annotated data and the LLM-generated ones (in different criteria), for model training. This should be alleviated by considering the near-distribution signals for direct utilized impact from the generated data samples~\cite{peng2024text,dietz2025principles}.
Thus, an effective framework for augmenting conversational dense retrieval should mix multiple principles, including multi-aspect augmentation, scalable generation, quality control, and near-distribution supervision to apply to conversational context-dependent queries and their associated relevance judgments.

To address the data scarcity issue and the difficulty of data augmentation for conversational dense retrieval, we propose \textbf{ConvMix}, a mixed-criteria framework to augment conversational dense retrieval, whose supported criteria are overlooked in previous studies, as shown in Table~\ref{table:comparison}.
Since generating the search-oriented conversational session and its corresponding relevant documents simultaneously is infeasible for LLMs~\cite{clarke2024llm,balog2025rankers}, we design a bidirectional relevance judgment augmentation from both query and document aspects: (i) generating augmented data from the existing manual relevance judgments by reformulating the corresponding conversational context-dependent queries; and (ii) generating pseudo generative relevance feedback~\citep{mackie2023generative} by rewriting the relevant documents in terms of associated queries.
This is achieved by deploying a generative LLM for scalable generation.
Then, a quality control mechanism is designed to select qualified samples based on the aspects of semantic diversity. 
Besides, a near-distribution supervision mechanism is applied to select the model parameter-sensitive samples from the bidirectional augmented and the original data based on utilization estimation.

With the augmented data via scalable generation and quality selection from our ConvMix framework, we leverage these generated data with the original ones to train conversational dense retrievers and evaluate on five widely used conversational search benchmarks. The experimental results show that our ConvMix outperforms previous strong baselines with various techniques and demonstrate the superior effectiveness of our proposed framework. 

\begin{table}[t]
    \centering
    \scalebox{0.8}{
    \begin{tabular}{lcccc}
    \toprule
    \small\multirow{2}{*}{\textbf{System}} &
    \small\textbf{Multi-aspect} & \small\textbf{LLM for} & \small\textbf{Quality} & \small\textbf{Near-Dist.} \\ 
    & \small\textbf{Augmentation} & \small\textbf{Scalable} & \small\textbf{Control} & \small\textbf{Supervision}\\
    \midrule
    DialogInpaint & \cmark & \xmark & \xmark & \xmark \\
    Converser & \xmark & \cmark & \xmark & \xmark \\ 
    ConvTrans & \xmark & \xmark & \xmark & \xmark\\
    ConvSDG & \cmark & \cmark & \xmark & \xmark\\
    ConvAUG & \xmark & \cmark & \cmark & \xmark\\
    \midrule
    ConvMix (Ours) & \cmark & \cmark & \cmark & \cmark\\
    \bottomrule
    \end{tabular}}
    \caption{The contained criteria comparison between ours and existing systems DialogInpaint~\cite{dai2022painting}, Converser~\cite{huang2023converser}, ConvTrans~\cite{mao2022convtrans}, ConvSDG~\cite{mo2024convsdg}, ConvAUG~\cite{chen2024generalizing}, including whether support for various criteria.}
    \label{table:comparison}
\end{table}

Our contributions are summarized as follows:

(1) We propose a mixed-criteria data augmentation framework for conversational dense, which supports scalable multi-aspect augmentation by reusing existing conversational session data and relevance judgments from query and document sides based on LLMs.

(2) We design a quality control mechanism to select qualified samples from the aspects of increasing the diversification of generated content and a near-distribution supervision mechanism to select the model parameter-sensitive sample based on utilization estimation.

(3) We conduct experiments on five widely used conversational search benchmarks with different settings and demonstrate the effectiveness of our approach by outperforming existing baseline methods.

\section{Related Work}
\subsection{Conversational Search}
Two main research lines in the literature are conducted for conversational search: (i) conversational query rewriting (CQR) and (ii) building a conversational dense retriever (CDR).
The CQR methods aim to convert the context-dependent queries into stand-alone ones and reuse any existing ad-hoc search retrievers~\cite{elgohary2019can,meng2025bridging}.  
Earlier CQR studies attempted to select useful tokens from the conversation context~\cite{voskarides2020query,fang2022open} or train a query rewriter with conversational sessions to mimic the human rewrites~\cite{yu2020few,lin2020conversational,vakulenko2021question}.
To mitigate the discrepancy between rewriter training and the downstream search task, some studies adopt reinforcement learning~\cite{wu2022conqrr,chen2022RLCQR} or exploit the ranking signals with the rewriting model training~\cite{mo2023convgqr,mao2023search,mo2023learning,yoon2024ask}.
Recent LLM-based methods generate the query directly rewrites for retrieval~\cite{ye2023enhancing,mao2023large,mo2024leverage,mo2026leveraging} or to generate high-quality pseudo queries as supervision signals for rewrite model training~\cite{jang2023itercqr,mo2024chiq,lai2025adacqr}.
The CDR methods~\cite{yu2021few,lin2021contextualized,mao2022curriculum,jin2023instructor,cheng2024interpreting,mao2024chatretriever,lupart2024disco,cheng2024coral,mo2025uniconv} directly encode the whole conversational search session into the model to perform end-to-end dense retrieval~\cite{kim2022saving,mo2024history}.
This paradigm highly relies on the quantity and quality of the training samples with relevance judgment, and a series of studies aim to augment the training data from different aspects~\cite{dai2022painting,mao2022convtrans,huang2023converser,mo2024convsdg,chen2024generalizing}. 

\subsection{Data Augmentation in Dense Retrieval}
The quantity and quality of data hold significant value for dense retrieval.
The development of LLMs provides the opportunity for scalable approaches to generate data for ad-hoc search models:
to generate queries from a document~\cite{arxiv22_InPars,ICLR23_Promptagator}, to generate a document from a query~\cite{gao2023precise,mackie2023generative,arxiv23_query2doc}, etc. 
In the context of conversational search, the data generation framework should also consider the consistency of the conversational sessions~\cite{joko2024doing,li2025personalized}. 
In addition, a crucial point is to consider the data distribution between the original data sample and the augmented parts from different annotated criteria~\cite {mo2024convsdg}, which requires avoiding generating homogeneous data and obtaining near-distribution supervision~\cite{peng2024text}. Besides, applying quality control mechanisms to select qualified samples is effective to reduce noise~\cite{chen2024generalizing}.
Our proposed ConvMix is a mixed-criteria framework that addresses and covers all these aspects, which are overlooked in previous studies.
\section{Motivation}

\subsection{Task Formulation}
Conversational search systems aim to identify relevant documents $d^+$ from a large collection $\mathcal{C}$ in response to the current query ($n$-th) $q_n$. 
This process is based on the context provided by the ongoing conversation session $\mathcal{X}=\{q_i\}_{i=1}^{n-1}$. Each query turn within a session depends on the preceding context, necessitating the conversational retriever to possess the capability to comprehend the user's search intent. Thus, having access to high-quality and adequate conversational search session data is important for model training.
The goal of augmenting conversational dense retrieval is to generate new session data $\mathcal{X^{\text{aug}}}=\{q_i^{\prime}\}_{i=1}^{n}$ with corresponding relevant query-document pairs $\{(q_i^{\prime}, d_i^{\prime})\}_{i=1}^{n}$.
Then, the combined annotated data $\{\mathcal{X}\} \cup \{\mathcal{X}^{\text{aug}}\}$ from existing data $\{\mathcal{X}\}$ and the automatically augmented data $\{\mathcal{X}^{\text{aug}}\}$ could be used for fine-tuning conversational dense retrieval.

\subsection{Diversity-Augmented Conversational Sessions}
We conduct preliminary experiments to investigate whether augmenting the data from existing annotated datasets by increasing the diversity of conversational sessions can impact the dense retrieval performance. 
The principle is motivated by data augmentation in the field of computer vision, which involves rotating the original photo~\cite{mikolajczyk2018data}.
In a similar pattern, we treat each conversational session as a photo, and achieving rotation for data augmentation by reordering the query turns in topic level.
Concretely, given an existing conversational session $s = \{{\mathcal{T}}_{1}, {\mathcal{T}}_{2}, \dots, {\mathcal{T}}_{t}\}$ with $t$ topics, where each topic $\mathcal{T}_i \in s$ would contain several query turns $q_i$, we apply random shuffle operation $\pi(\cdot)$ as a permutation function to the topic order in this conversational session to obtain a new one $s^{\prime}$, i.e., $s = \pi(s^{\prime})$. 
As a result, the new session $s^{\prime} = \{{\mathcal{T}}_{\pi(1)}, {\mathcal{T}}_{\pi(2)}, \dots, {\mathcal{T}}_{\pi(t)}\}, \pi\neq \text{id}$, is a permutation of the original session $s$, i.e., they are content-equivalent but order-different.
By applying this operation on all conversational sessions $\mathcal{X}$ in the original dataset, we obtain double the data $\mathcal{X}_\text{DA}^{\text{aug}}=\{\mathcal{X}\} \cup \{\mathcal{X}_\text{DA}^\prime\}$ by augmenting from the content diversification aspect. 
Then, we conduct conversational dense retriever fine-tuning with augmented data $\mathcal{X}_\text{DA}^{\text{aug}}$ and dense model ANCE~\cite{xiong2020approximate}, which is evaluated on two datasets with available topic information, TopiOCQA~\cite{adlakha2022topiocqa} and QReCC~\cite{anantha2021open}.
The results are shown in Table~\ref{table: preliminary}, where we can observe that augmenting existing datasets by increasing the diversity of conversational sessions can improve the performance of the conversational dense retriever.
With the validated conjecture, we continue to investigate the augmenting conversational dense retrieval by increasing content diversification and achieving quality control with mixed-criteria.

\begin{table}[h]
\centering
\begin{tabular}{lcccc}
\toprule
& \multicolumn{2}{c}{TopiOCQA} & \multicolumn{2}{c}{QReCC} \\
\cmidrule(lr){2-3}\cmidrule(lr){4-5}
~ & {MRR} & {R@10} & {MRR} & {R@10} \\
\midrule
Original & 22.9 & 43.0 & 46.3 & 70.4 \\
+ Augmentation & 29.8 & 49.2 & 49.1 & 72.5\\
\bottomrule
\end{tabular}
\caption{Preliminary experiments on diversity-oriented augmented conversational search data on two datasets.}
\label{table: preliminary}
\end{table}

\section{Methodology}

\subsection{Overview}
\begin{figure*}
    \centering
    \includegraphics[width=0.9\linewidth]{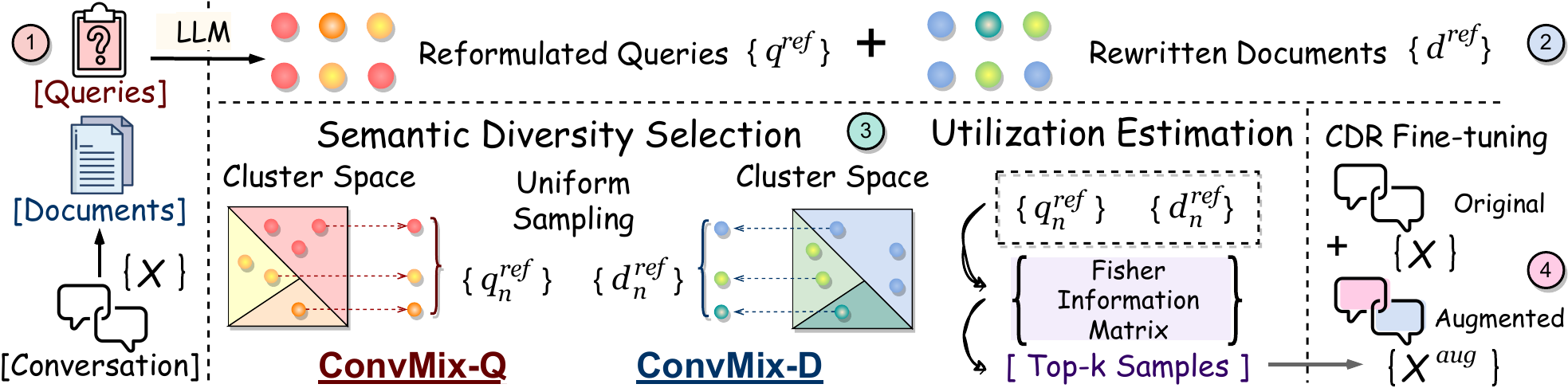}
    \caption{Illustration of our ConvMix framework, including bidirectional augmentation, two quality control mechanisms with semantic and utilization selection, and mixing original and augmented data for conversational dense retrieval fine-tuning.}
    \label{fig:method}
\end{figure*}

Augmenting conversational dense retrieval training data with multiple criteria is a non-trivial task.
The augmented data generation should be aware of various aspects, including multi-aspect augmentation~\cite{balog2025rankers} in a scalable way~\cite{mo2024convsdg}, data quality control by filtering noisy samples~\cite{chen2024generalizing}, and obtain near-distribution supervision to fit different types of annotated data~\cite{peng2024text}.
By considering all these aspects, we propose \textbf{ConvMix}, a mixed-criteria framework to augment conversational dense retrieval data, whose overview is presented in Figure~\ref{fig:method}.
Several components are integrated in our \textbf{ConvMix}, including (i) augmenting relevance judgment from two-sides by reformulating context-dependent query and rewriting its  relevant documents, (ii) selecting semantic diversity samples by clustering to increase the diversification of generated data content, and (iii) selecting model parameter-sensitive sample with near-distribution supervision by utilization estimation.

\subsection{Bidirectional Relevance Judgment Augmentation} 
One crucial criteria is to conduct data augmentation from multiple aspects to increase the diversification of the generated content in a scalable manner.
However, even with the advanced LLMs, it is still infeasible to generate the search-oriented conversational session and its corresponding relevant documents simultaneously~\cite{clarke2024llm,sekulic2024analysing} and increasing the diversity of the generated results~\cite{peng2024text}.
To reduce the difficulty for LLMs and fully leverage the existing annotated datasets, we conduct data augmentation from two directions to separate the generation of conversational sessions and relevance judgment annotations, which can also generate data from multiple aspects to increase the diversity.

On one hand, we augment the data from the query side as \textbf{ConvMix-Q} variant, which aims to reuse the annotated relevance judgments in existing datasets by reformulating the context-dependent queries. 
We expect the reformulated query turns $q_n^{\text{ref}}$ into another expression could keep the same search intent as the original one $q_n$, and then enhance the diversity of the generated new session after applying on every query turn.
On the other hand, we conduct data augmentation from the existing relevant documents side as \textbf{ConvMix-D} variant, where we rewrite each relevant document $d_n^+$ for its associated query turn $q_n$ and its conversational context $\{q_i\}_{i=1}^{n-1}$.
The rewritten relevant documents $d_n^{\text{ref}}$ serve as generative relevance feedback~\cite{mackie2023generative}, which do not exist in the collection but share the same semantic information as the original documents.
Although the rewritten relevant documents might contain wrong information or hallucination, it can still benefit to learn the semantic representation match for retriever model training~\cite{arxiv22_InPars,peng2024text}.
Eventually, the relevance judgments are augmented from both sides based on the existing datasets, which can be formulated as 
$$
    q_n^{\text{ref}} = \mathcal{LLM}(q_n | \{q_i\}_{i=1}^{n-1}), \ d_n^{\text{ref}} = \mathcal{LLM}(d_n^+ | \{q_i\}_{i=1}^{n-1}, q_n)
$$
where the ($\{q_i\}_{i=1}^{n-1} \circ q_n, d_n^+$) is the available relevance judgment in the existing datasets.

To scale up the size of the augmented data, an intuitive approach is to produce multiple reformulated queries or rewritten documents, i.e., $m$ variant of $q_n^{\text{ref}}$ and $d_n^{\text{ref}}$ would be generated as $\{q_n^{\text{ref}}\}_1^{m}$ and $\{d_n^{\text{ref}}\}_1^{m}$ ($m>1$), in \textbf{ConvMix-Q} and \textbf{ConvMix-D}, respectively.
The generated data from both sides corresponding to all turns consists of the augmented set $\mathcal{X}^{\text{aug}}$.
Then, the follow-up question is how to control the quality and select the utilized ones of the generated data.

\subsection{Clustering for Semantic Diversity Sample Selection}
The pseudo data generated by LLM might be homogeneous and thus using them for model training would result in overfitting and compromise the model's robustness~\cite{peng2024text}. 
Therefore, increasing the diversification of the generated sample is crucial to mitigate this issue~\cite{wang2025diversity}.
However, the common practice to improve the diversity of the generated content is to adjust the temperature hyperparameter within the generative models, which does not directly impact the semantic diversity.
To this end, we leverage clustering to select the generated data sample from the semantic diversity aspect.
Specifically, with $m$ generated content, e.g., reformulated queries $\{q_n^{\text{ref}}\}_1^{m}$ or rewritten relevant documents $\{d_n^{\text{ref}}\}_1^{m}$, for each ($n$-th) turn, a k-means clustering algorithm is used to partition them into $k$ disjoint clusters as
$$
\mathcal{C} = \{C_1, C_2, \ldots, C_k\}, \quad \bigcup_{i=1}^k C_i = \{q_n^{\text{ref}}\}_1^{m} \ \text{or} \ \{d_n^{\text{ref}}\}_1^{m}
$$
where $C_i \cap C_j = \emptyset \text{ for } i \ne j$. 

Then, from each cluster \( C_i \), one item is randomly sampled as $x_{\text{ni}}^{\text{Div}} \sim \text{Uniform}(C_i)$. The final output set \( x_{\text{n}}^{\text{Div}} \) $\subseteq \{q_n^{\text{ref}}\}_1^{m} / \{d_n^{\text{ref}}\}_1^{m}$ contains \( k \) augmented data samples and the whole procedure is formulated as 
$$
x_{\text{n}}^{\text{Div}} = \left\{ x_{\text{ni}}^{\text{Div}} \sim \text{Uniform}(C_i) \,\middle|\, C_i \in \text{Cluster}_k(\mathcal{X}^{\text{gen}}) \right\}
$$
where $\mathcal{X^{\text{gen}}}$ could be either $\{q_n^{\text{ref}}\}_1^{m} \ \text{or} \ \{d_n^{\text{ref}}\}_1^{m}$. Based on the hyperparameter $k$, we can obtain $k$-fold augmented data from both the reformulated query $q_n^{\text{ref}}$ and the rewritten relevant document $d_n^{\text{ref}}$ sides.

\subsection{Utilization Estimation for Near-Distribution Supervision Selection via Fisher Information}
Mixing data with various annotated types, i.e., the bidirectional augmented data and the original data, for model training might result in data distribution drift~\cite{guilhaumon2024data,zhang2025batch}.
To mitigate this potential issue, the crucial solution is to obtain near-distribution supervision~\cite{peng2024text} by selecting those samples that are ``most useful for model update'' to ensure that the training with augmented data is both efficient and less noisy, and better impact on the utilized fitting of model parameters~\cite{liu2020quantum}.
Intuitively, a data sample that is highly sensitive to the model's output is considered to carry significant information and thus such a sample is closer to the ideal data distribution and should thus be selected.

To achieve it, we leverage Fisher Information Matrix (FIM)~\cite{fisher1922mathematical} to measure the sensitivity of model parameter estimation in terms of a given augmented data sample from the previous selected set $\{\mathcal{X}^{\text{Div}}\}$.
As shown in Algorithm~\ref{alg: fisher}, we compute the FIM score for each pseudo-labeled relevant query-document pair ($q_n^{\text{ref}},d_n^{\text{ref}}$) by estimating the squared gradient norm of the prediction loss based on an ad-hoc search model used for conversational dense retrieval fine-tuning later. The top-$k$ pairs with the highest scores are retained for downstream training, as they are assumed to provide the most informative updates to the model parameters. 
Eventually, the final mixed augmented datasets with near-distribution supervision are used for conversational dense retriever training as a combination \textbf{ConvMix-Combine}.

\begin{algorithm}[h]
\caption{Utilized Selection via Fisher Information}
\label{alg: fisher}
\begin{algorithmic}[1]
\REQUIRE Pseudo-labeled query-document pairs $\mathcal{X}_n^{\text{Div}} = \{(q_1^{\text{ref}}, d_1^{\text{ref}}), \dots, (q_m^{\text{ref}}, d_m^{\text{ref}})\}$ by data augmentation for a query turn $q_n$, retrieval model $f(\cdot,\cdot;\theta)$, top-$k$ value $k$
\ENSURE Selected final augmented set $\mathcal{X}_n^{\text{aug}}$
\STATE Initialize empty list $\mathcal{L} \leftarrow [\,]$
\FOR{each $(q_i^{\text{ref}}, d_i^{\text{ref}}) \in \mathcal{X}_n^{\text{Div}}$}
    \STATE Compute similarity score $s_i = f(q_i^{\text{ref}}, d_i^{\text{ref}}; \theta)$
    \STATE Set retrieval score $r_i = f(q_i, d_i^{\text{ref}}; \theta)\ \text{or} \ f(q_i^{\text{ref}}, d_i^+; \theta) $
    \STATE Compute loss $\ell_i = \text{Loss}(s_i, r_i)$
    \STATE Compute gradient $\nabla_\theta \ell_i$ via backpropagation
    \STATE Estimate Fisher information $F_i = \|\nabla_\theta \ell_i\|^2$
    \STATE Append $(q_i^{\text{ref}}, d_i^{\text{ref}}, F_i)$ to $\mathcal{L}$
\ENDFOR
\STATE Sort $\mathcal{L}$ in descending order by $F_i$
\STATE Select top $k$ pairs to form $\mathcal{X}_n^{\text{aug}}$
\RETURN $\mathcal{X}_n^{\text{aug}}$
\end{algorithmic}
\end{algorithm}

\subsection{Conversational Dense Retrieval Fine-Tuning}
With the augmented data $\{\mathcal{X}^\text{aug}\}$, we perform conversational dense retrieval fine-tuning together with the original data $\{\mathcal{X}\}$ in a semi-supervised learning paradigm. ``Semi-supervised'' means we combine the original data points with manual labels and the generated data points without manual labels for model training, which enables the augmented parts with consistency regularization.
We employ a widely used ad-hoc search retriever, ANCE~\cite{xiong2020approximate}, which serves as both the query and document encoder, denoted as $\mathcal{F}_Q$ and $\mathcal{F}_D$, respectively.
In this process, we consider all preceding queries within the same session to concatenate with the current query turn as the session representation $q_n^{\text{concat}}$ as
\begin{equation}
    q_n^{\text{concat}} = q_1 \circ \cdots q_{i} \cdots \circ q_{n-1} \circ q_n, \quad i \in [1, n-1]
    \label{eq: query form}
\end{equation}
where $\circ$ denotes concatenation. Then, a similarity function $\mathcal{S}$ based on dot product is applied to score a candidate document $d$ as Eq.~\ref{eq: similarity}. During the training phase, we update only the query encoder, while the document encoder is frozen. We follow the simplest fine-tuning principle with only in-batch negatives, which aims to directly reflect the impact of the augmented data rather than involving other advanced techniques in fine-tuning, e.g., hard negatives mining~\cite{kim2022saving} or injecting other signals~\cite{mo2024aligning}. 
\begin{equation}
    \mathcal{S}(q_n^{\text{concat}}, d) = \mathcal{F}_{Q}(q_n^{\text{concat}})^{T} \cdot \mathcal{F}_{D}(d)
    \label{eq: similarity}
\end{equation}
The final training objective follows the widely used contrastive learning loss as
\begin{equation}
\mathcal{L} = \frac{e^{\mathcal{S}(q_n^{\text{concat}}, d^+)}} {e^{\mathcal{S}(q_n^{\text{concat}}, d^+)} + \sum_{d_n^- \in \mathcal{C}^-} e^{\mathcal{S}(q_n^{\text{concat}}, d^-)}}
    \label{eq: loss}
\end{equation}
where the $d^+$ and $d^-$ denote the positive and negative samples for each query turn. During the inference, we perform the Approximate Nearest Neighbors search based on the dense index using Faiss~\citep{johnson2019billion}.

\section{Experiments}
\subsection{Experimental Setup}
\noindent \textbf{Datasets and Evaluation Metric.}
We use five conversational search datasets to evaluate our methods in different settings. All datasets contain relevance judgments for each original query, and the detailed statistical information is provided in the Appendix.
The TopiOCQA~\cite{adlakha2022topiocqa} and QReCC~\cite{anantha2021open} datasets are used to produce augmented data on top of their original relevance judgment annotations and conversational sessions.
Then, the original data and the augmented data are combined to train conversational dense retrievers for the usual training-test evaluation. 
The three 
datasets~\cite{dalton2020trec,dalton2021cast,dalton2022cast} with only a few samples are used for out-of-domain evaluation (i.e., directly apply the models trained on TopiOCQA to the CAsT datasets). 
For a thorough comparison with existing systems, we use four standard metrics: MRR, NDCG@3, Recall@10, and Recall@100 to evaluate the retrieval effectiveness. Significance tests are conducted using paired t-tests at $p < 0.05$ level compared to the main competitors.
We adopt the \textit{pytrec\_eval} tool~\cite{sigir18_pytrec_eval} for metric computation.

\begin{table*}[t]
    \centering
    \begin{tabular}{c|l|ccc|ccc}
    \toprule
    \multirow{2}{*}{Category} & \multicolumn{1}{c|}{\multirow{2}{*}{{Method}}} & \multicolumn{3}{c|}{TopiOCQA} &
    \multicolumn{3}{c}{QReCC}\\
    \cmidrule(lr){3-5} \cmidrule(lr){6-8}
     & & {MRR} & {NDCG@3} & {Recall@10} & {MRR} & {NDCG@3} & {Recall@10} \\
    \midrule
    \multirow{5}{*}{Advanced CDR} & InstructorR & 25.3 & 23.7 & 45.1  & 43.5 & 40.5 & 66.7 \\
    & SDRConv & 26.1 & 25.4 & 44.4 &  47.3 & 43.6 & 69.8 \\
    & ConvDR & 27.2 & 26.4 & 43.5 & 38.5 & 35.7 & 58.2 \\
    & HAConvDR & 30.1 & 28.5 & 50.8 & 48.5 & 45.6 & 72.4 \\
    & QRACDR & 31.8 & 30.6 & 50.0 & 47.1 & 44.5 & 71.4 \\
    \midrule 
    \multirow{6}{*}{Augmented CDR} & ConvTrans & 28.6 & 27.4 & 48.1 & 48.2 & 46.9 & 70.6 \\
    ~ & ConvSDG & 28.4 & 27.0 & 47.9 & 49.2 & 47.7 & 73.4 \\
    ~ & ConvAUG & 35.0 & 33.3 & 57.9 & 52.7 & 50.4 & \textbf{75.6}\\
    \cmidrule(lr){2-8}
    ~ & ConvMix-Q & 36.9 & 35.8 & 57.9 & 51.3 & 49.8 & 74.7\\
    ~ & ConvMix-D & 34.4 & 33.3 & 55.2 & 50.8 & 49.2 & 74.0\\
    ~ & ConvMix-Combine & \textbf{37.7} & \textbf{36.6} & \textbf{58.7} & \textbf{52.9} & \textbf{50.7} & \textbf{75.6}\\
    \bottomrule
    \end{tabular}
    \caption{Performance of different dense retrieval methods on two datasets. The number of training samples for our ConvMix-Q and ConvMix-D is double that of the original datasets, as the methods in Advanced CDR.
    The training sample for the other three main competitors, ConvTrans, ConvSDG, and ConvAUG, is the same as shown in Table~\ref{table: Zero-shot}. \textbf{Bold} indicates the best results.}
    \label{table: Main Results}
\end{table*}

\begin{table*}[t]
    \centering
    \begin{tabular}{l|c|ccc|ccc|ccc}
    \toprule
        \multicolumn{1}{c|}{\multirow{2}{*}{Method}} & \multirow{2}{*}{Training Sample} & \multicolumn{3}{c|}{CAsT-19} & \multicolumn{3}{c|}{CAsT-20} & \multicolumn{3}{c}{CAsT-21} \\ 
        \cmidrule(lr){3-5} \cmidrule(lr){6-8} \cmidrule(lr){9-11} 
        & & MRR     & N@3   & R@100  & MRR     & N@3   & R@100  & MRR     & N@3   & R@100  \\ 
        \midrule
        No Augmentation & 45,450 & 66.2 & 40.1 & 29.2 & 42.5 & 26.5 & 31.6 & 46.7 & 29.1 & 37.4\\
        ConvTrans & 408,389 & \textbf{73.2} & \textbf{45.3} & 36.0 & \underline{45.9} & 31.2 & 38.7 & 47.9 & 35.2 & 38.8\\
        ConvAUG & 136,858 & - & - & - & 45.0 & 30.7 & 38.4 & 54.8 & 36.8 & 45.9 \\
        ConvSDG & 89,400 & 63.2 & 37.2 & \textbf{36.9} & 40.4 & 29.7 & 36.5 & 49.3 & 33.4 & 38.2 \\
        \midrule
        ConvMix-Q & 90,900 & 64.5 & 37.8 & 36.6 & 45.4 & \underline{32.6} & 39.1 & \underline{49.9} & 34.6 & 42.5\\
        ConvMix-D & 90,900 & 67.4 & 41.4 & 36.3 & 44.7 & 31.5 & 38.3 & \underline{58.3} & \underline{37.6} & \underline{46.3}\\
        ConvMix-Combine & 136,350 & \underline{68.2} & \underline{44.3} & \underline{36.8} & \textbf{46.1} & \textbf{33.0} & \underline{39.7} & \textbf{60.3} & \textbf{38.4} & \textbf{47.2} \\
        \bottomrule
     \end{tabular}
     \caption{Out-of-Domain performance on three CAsT datasets. \textbf{Bold} and \underline{underline} indicate the best and the second-best results.}
     \vspace{-2ex}
     \label{table: Zero-shot}
\end{table*}

\noindent \textbf{Baselines Methods.}
We compare our method with two categories of baselines: data augmentation methods for conversational dense retrieval (Augmented CDR), and conversational dense retrieval approaches with advanced fine-tuning techniques (Advanced CDR).
The first category serves as main competitors, which includes the following systems: (1) \textbf{ConvTrans}~\cite{mao2022convtrans}: An approach to transform session search data into conversational search data for data augmentation. (2) \textbf{ConvSDG}~\cite{mo2024convsdg}: A data augmentation approach to generate session-level conversational sessions with pseudo-relevant feedback as relevance judgments, and (3) \textbf{ConvAUG}~\cite{chen2024generalizing}: a framework to generate multi-level augmented conversations to capture the diverse nature of conversational contexts via LLM with human cognition mechanism.
The second line of advanced CDR includes previous systems: (4) \textbf{InstructorR}~\cite{jin2023instructor}: An unsupervised conversational dense retrieval method by instructing LLMs to produce LLM-based representation of the documents to serve as additional features. (5) \textbf{SDRConv}~\cite{kim2022saving}: a fine-tuning method with different mined hard negatives. (6) \textbf{ConvDR}~\cite{yu2021few}: A method to leverage both conversational search data and manually rewritten queries as supervision signals for conversational dense retrieval fine-tuning. (7) \textbf{HAConvDR}~\cite{mo2024history}: An approach fine-tuned on context-denoising reformulated query and additional signals from historical turns based on the impact of retrieval effectiveness, and (8) \textbf{QRACDR}~\cite{mo2024aligning}: A query representation alignment conversational dense retriever to align the query representation with those of rewritten queries and relevant documents.

\noindent \textbf{Implementation Details.}
We implement the conversational dense retriever based on ANCE~\cite{xiong2020approximate} using PyTorch and Huggingface libraries, ensuring all baseline systems are in a comparable setting. 
For data augmentation, each query turn is reformulated and relevant documents are rewritten on Mistral-7B-Instruct-v0.3~\cite{jiang2023mistral7b}, whose instruction templates are provided in the Appendix.
The scalable relevance judgment augmentation is investigated from 1-fold to 10-fold.
The top-$k$ sample selection for semantic diversity clustering and the FIM score for near-distribution are both set to 3.
For conversational dense retrieval fine-tuning, following previous studies~\cite{yu2021few,mao2023learning}, the lengths of the query turn, session query, and document are truncated into 64, 512, and 384, respectively, to fit the majority of examples in the dataset. The batch size is set to 32 in accordance to our computational resources. 
We use Adam optimizer with 1e-5 learning rate and set the training epoch to 10. All experiments are conducted on four Nvidia V100 32G GPUs. More details are provided in our code repository: \url{https://github.com/fengranMark/ConvMix}.

\begin{table*}[t]
\centering
\begin{tabular}{l|cc|cc|cc|cc|cc}
\toprule
  {\multirow{2}{*}{Ablation}}& \multicolumn{2}{c|}{TopiOCQA} & \multicolumn{2}{c|}{QReCC} & \multicolumn{2}{c|}{CAsT-19} & \multicolumn{2}{c|}{CAsT-20} & \multicolumn{2}{c}{CAsT-21}\\
\cmidrule(lr){2-3}\cmidrule(lr){4-5}\cmidrule(lr){6-7}\cmidrule(lr){8-9}\cmidrule(lr){10-11}
~ & {MRR} & {N@3} & {MRR} & {N@3} & {MRR} & {N@3} & {MRR} & {N@3} & {MRR} & {N@3}\\
\midrule
Full Model (ConvMix-Combine) & 37.7 & 36.6 & 52.9 & 50.7 & 68.2 & 44.3 & 46.1 & 33.0 & 60.3 & 38.4 \\
\quad w/o Semantic Selection & 36.2 & 35.1 & 50.6 & 48.9 & 66.4 & 40.2 & 45.2 & 31.9 & 57.8 & 37.0\\
\quad w/o Utilization Selection & 33.0 & 32.0 & 47.5 & 46.8 & 63.7 & 36.5 & 42.4 & 29.8 & 54.2 & 35.9\\
\quad w/o Query-side Augmentation & 34.4 & 33.3 & 50.8 & 49.2 & 67.4 & 41.4 & 44.7 & 31.5 & 58.3 & 37.6\\
\quad w/o Doc-side Augmentation & 36.9 & 35.8 & 51.3 & 49.8 & 64.5 & 37.8 & 45.4 & 32.6 & 49.9 & 34.6\\
\quad w/o Original Data & 32.2 & 32.9 & 46.2 & 45.3 & 60.1 & 35.1 & 36.7 & 25.1 & 48.0 & 31.1\\
\bottomrule
\end{tabular}
\caption{Ablation studies on five datasets within various components of our ConvMix.}
\vspace{-2ex}
\label{table: ablation study}
\end{table*}

\subsection{Overall Performance}
We first conduct the standard training-testing evaluation on TopiOCQA and QReCC, combining with the augmented data based on the original dataset. 

The overall performance is shown in Table~\ref{table: Main Results}. 
We can observe that our ConvMix-Combine with data augmentation from combining reformulating query turns and rewriting relevant documents achieves the best performance on both datasets. 
Specifically, our ConvMix-Combine outperforms the main competitors ConvTrans, ConvSDG, and ConvAUG with 9.1\%, 9.2\%, and 3.3\% absolute gains, respectively, on TopiOCQA in terms of MRR score, which demonstrates the effectiveness of augmenting the existing dataset from various aspects.
The superior performance might be attributed to the quality control from the aspects of increasing semantic diversity and utilization estimation via clustering and FIM scores.
The ConvMix-Q variant is consistently better than ConvMix-D on two datasets, and they also outperform existing methods except for comparing to ConvAUG on QReCC, which indicates the importance of quality control mechanisms in ConvAUG and our ConvMix, while lacking in previous augmented CDR methods.
Besides, we can find that ConvAUG and our ConvMix perform better than the previous advanced CDR methods, which emphasizes the potential of augmenting conversational dense retrieval via existing datasets.
Further investigation could find a better way to connect augmenting CDR with advanced fine-tuned techniques.

\subsection{Out-of-Domain Performance}
We then conduct out-of-domain testing on three CAsT datasets to evaluate the robustness of our ConvMix. The results are shown in Table~\ref{table: Zero-shot}, where the used training sample for each compared baseline is reported.

We can see that our ConvMix outperforms previous methods on two newer and more challenge datasets, CAsT-20 and CAsT-21, and is on par with those that have similar training samples on CAsT-19. 
The better performance achieved by ConvMix with less training samples, which demonstrate its efficiency in data usage and the higher generated data in our mix-criteria framework.
Besides, we observe similar results that better performance could be obtained by combining reformulating query turns and rewriting relevant documents, i.e., ConvMix-Combine performs better than using either ConvMix-Q or ConvMix-D alone, which indicates our data augmentation framework can produce more diverse generated results from various aspects and effectively to mix them via near-distribution supervision.

\subsection{Ablation Studies}
In this section, we conduct ablation studies to investigate the effects of incorporating various components into our ConvMix framework, from the aspects of the quality control selection and the used annotated data types. The results are reported in Table~\ref{table: ablation study}.

In terms of the quality control mechanism, the semantic diversity selection and the utilization estimation selection are both effective across all datasets, indicating again the necessity of selecting high-quality samples in the data augmentation schema.
The utilization estimation selection contributes more than the semantic diversity selection. The reason might be that the utilization estimation for near-distribution supervision based on the FIM information can more directly impact the model parameter updating and better fit the ideal data distribution than enhancing the semantic diversity.

In the context of the usage of various types of annotated data, we can observe that the original data is important to prevent data distribution drift, while using the augmented data alone can also achieve a competitive performance. 
Besides, the augmentation from the query and document sides are both effective. They can complement each other and further boost the performance with a combination. This phenomenon reveals that our ConvMix is robust to mix various annotated data and fully exploits their contribution for obtaining clean and near-distribution augmented data as supervision signals.

\subsection{Ratio of Augmented Training Samples Impact}
\begin{figure}[h]
    \centering  \includegraphics[width=\linewidth]{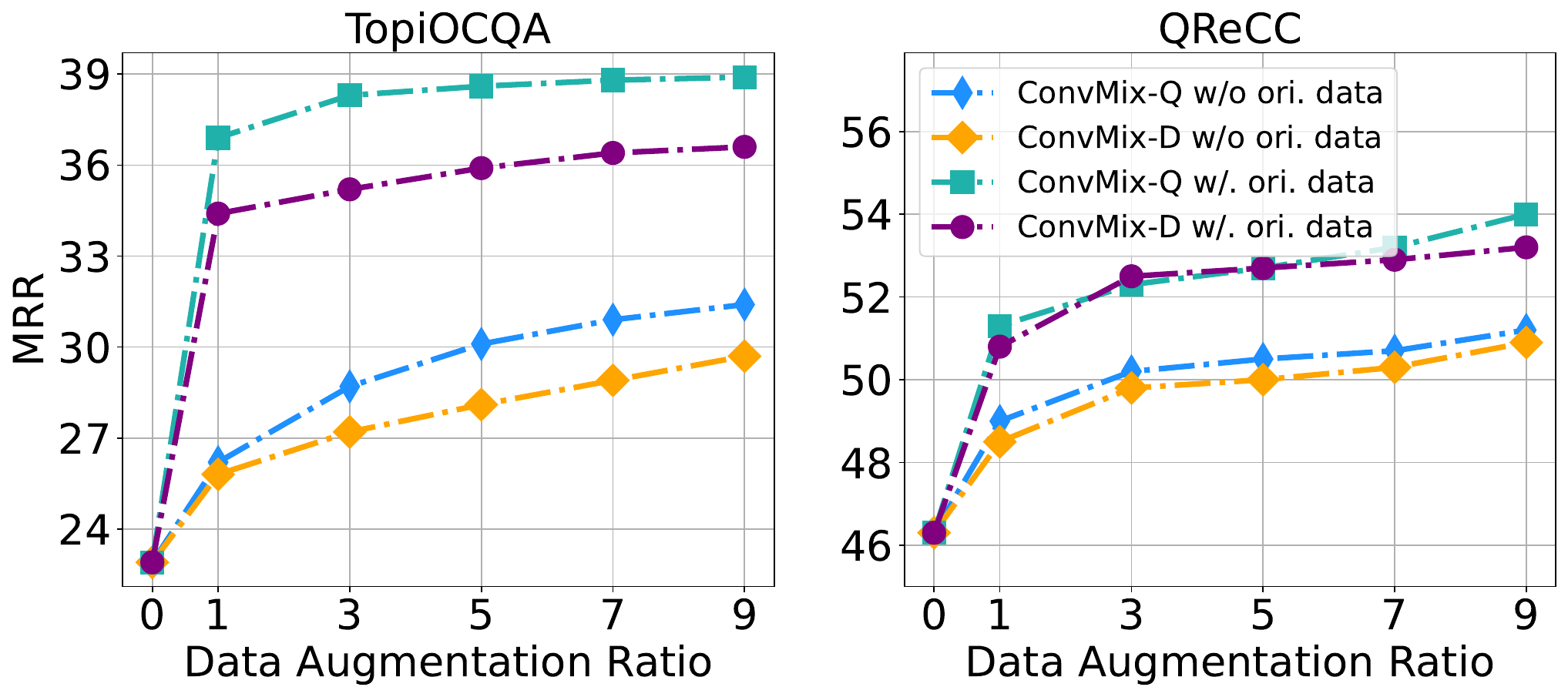}
    \caption{Model performance at the MRR score with various ratios of augmented training samples on two datasets.}
    \vspace{-2ex}
    \label{fig: augmentation}
\end{figure}

To study the impact of the amount of training data, we test the performance of two variants of our ConvMix with different augmented ratios. The results are shown in Figure~\ref{fig: augmentation}.

We find that by increasing the augmentation scale of the generated training data, the retrieval performance shows an upward trend, but gradually becomes slower. 
The observation indicates that although increasing the amount of augmented training data can improve the effectiveness of the retrieval model, simply increasing it cannot achieve better results.
This is because the noise and homogeneous information are also increasing among the generated data.
Thus, it is necessary to improve the data quality for conversational dense retrieval by quality selection mechanisms, which demonstrate the importance of our designed approaches via increasing semantic diversity and estimating utilization scores for high-quality sample selection.

Besides, we can also observe that the effectiveness of data augmentation is more obvious on the TopiOCQA dataset than on QReCC. This might be related to the original construction of the dataset, where the TopiOCQA contains multiple topics within a conversational session than QReCC. Thus, the original TopiOCQA provides more abundant semantic information and makes it more possible to achieve effective data augmentation than QReCC. These phenomenon emphasizes the importance of content diversity in augmented data.

\section{Conclusion}
In this paper, we introduce \textbf{ConvMix}, a mixed-criteria framework to augment conversational dense retrieval with multi-aspect augmentation in a scalable manner. 
To ensure the highly-quality of the generated training sample, we improve the content diversification from semantic aspect. 
We further design a utilization estimation mechanism to obtain near-distribution supervision to combine various annotated data for effective conversational dense retrieval fine-tuning.
The experimental results on five widely used conversational search benchmarks demonstrate the effectiveness of our framework by outperforming different existing baseline methods.
Our study shows the potential of using an LLM to generate additional training data with multiple criteria for fine-tuning a dense retriever, which enriches the already extensive body of studies trying to exploit LLMs for search. More research is still required to find the best way to leverage LLMs for enhancing conversational search.

\bibliography{aaai2026}

\clearpage
\appendix
\section*{Technical Appendix}
\label{sec: technical_appendix}
\subsection{Prompt Template Used for Two-sided Relevance Judgments Augmentation}
We provide our prompts for instructing LLMs to generate both diverse reformulated context-dependent queries and distinct generative relevance feedback on top of existing relevance judgments. The prompts are available in Table~\ref{tab: prompt_template}.

\section*{Code and Data Appendix}
\label{sec: Code_Data_Appendix}
\subsection{Datasets}
The five widely used datasets, TopiOCQA, QReCC, CAsT-19, CAsT-20, and CAsT-21, in our paper can be downloaded from the websites provided in the original paper.
The detailed statistical information is provided in Table~\ref{table: dataset}.
\begin{table}[h]
\centering
\setlength{\tabcolsep}{4pt}{
\begin{tabular}{llrrr}
\toprule
Dataset & Split & \#Conv. & \#Turns(Qry.) & \#Collection \\ \midrule
\multirow{2}{*}{TopiOCQA} & Train & 3,509 & 45,450 & \multirow{2}{*}{25M} \\
 & Test  & 205 & 2,514 & \\
\midrule
\multirow{2}{*}{QReCC} & Train & 10,823 & 29,596 & \multirow{2}{*}{54M} \\
 & Test  & 2,775 & 8,124 & \\
\midrule
CAsT-19 & Test & 50 & 479 & \multirow{2}{*}{38M}\\
CAsT-20 & Test & 25 & 208 & \\
\midrule
CAsT-21 & Test & 26 & 239 & 40M \\
\bottomrule
\end{tabular}}
\caption{Statistics of conversational search datasets.}
\label{table: dataset}
\end{table}

\begin{table}[t]
\centering
\begin{tabular}{p{\columnwidth}}
\toprule
    \multicolumn{1}{c}{\textbf{Prompt for reformulating context-dependent queries}}\\
\midrule
\# You will be given a user question. Please provide \{k\} equivalent questions, such that each of the \{k\} questions has the same meaning but is in a different form. Write each query on one line.\\
\\
\# Here is the User Question:\\
\\
\{A context-dependent user query\}\\
\\
\# And its associated conversational context:\\
\\
\{A conversational session\}\\
\\
\# Now give me the \{k\} different questions. Don't say any other words. Don't generate a sequence number of indicators. Just write each question on one line.\\

\midrule
    \multicolumn{1}{c}{\textbf{Prompt for generating pseudo relevance feedback}}\\
\midrule
\# Rewrite the following document. While keeping the entities, proper nouns, and key details such as names, locations, and terminology intact, create \{k\} new versions of the text that express the same ideas in a different way. Make sure the revised documents are all distinct from the original one, but preserve the core meaning and relevant information. Here is the document to be rewritten:\\
\\
\{A reference document annotated as relevant to a given query\}\\
\\
\# Now give me the \{k\} different and informative revised documents. The format should be:\\
\\
document1\\
document2\\
document3\\
....\\
document10.\\
\\
Each document should be on one line.\\

\bottomrule
\hline
\end{tabular}
\caption{Prompt Template for instructing LLMs to generate augmented reformulated context-dependent queries and rewritten generative relevance feedback.} 
\label{tab: prompt_template}
\end{table}

\clearpage
\end{document}